\documentclass[sn-aps]{sn-jnl}

\usepackage{bm}
\usepackage{color}
\usepackage{txfonts}
\usepackage{amssymb, amsmath}
\usepackage{physics}
\usepackage{comment}

\jyear{2022}%

\theoremstyle{thmstyleone}%
\theoremstyle{thmstyletwo}%

\theoremstyle{thmstylethree}%

\begin{document}

\title[Gaussian breeding for encoding a qubit in propagating light]{\ \ \ \ \ \ \ Gaussian breeding for encoding\\ a qubit in propagating light}


\author*[1,2]{\ \ \ \ \ \ \ \ \fnm{Kan} \sur{Takase}}\email{takase@ap.t.u-tokyo.ac.jp}

\author[1]{\fnm{Kosuke} \sur{Fukui}}\email{fukui@alice.t.u-tokyo.ac.jp}

\author[1]{\fnm{Akito} \sur{Kawasaki}} \email{kawasaki@alice.t.u-tokyo.ac.jp}

\author[1,2]{\\ \fnm{Warit} \sur{Asavanant}}\email{warit@ap.t.u-tokyo.ac.jp}

\author[1,2]{\fnm{Mamoru} \sur{Endo}}\email{endo@ap.t.u-tokyo.ac.jp}

\author[2]{\fnm{Jun-ichi} \sur{Yoshikawa}}\email{jun-ichi.yoshikawa@riken.jp}

\author[3]{\\ \fnm{Peter} \spfx{van} \sur{Loock}}\email{loock@uni-mainz.de}

\author*[1,2]{\fnm{Akira} \sur{Furusawa}}\email{akiraf@ap.t.u-tokyo.ac.jp}

\affil[1]{\orgdiv{Department of Applied Physics, School of Engineering}, \orgname{The
University of Tokyo}, \orgaddress{\street{7-3-1, Hongo}, \city{Bunkyo}, \postcode{113-8656}, \state{Tokyo}, \country{Japan}}}

\affil[2]{\orgdiv{Optical Quantum Computing Research Team}, \orgname{RIKEN Center
for Quantum Computing}, \orgaddress{\street{2-1, Hirosawa}, \city{Wako}, \postcode{351-0198}, \state{Tokyo}, \country{Japan}}}


\affil[3]{\orgdiv{Institute of Physics}, \orgname{Johannes-Gutenberg University of Mainz}, \orgaddress{\street{7}, \city{Staudingerweg}, \postcode{55128}, \state{Mainz}, \country{Germany}}}




\abstract{\textbf{Practical quantum computing requires robust encoding of logical qubits in physical systems to protect fragile quantum information. Currently, the lack of scalability limits the logical encoding in most physical systems, and thus the high scalability of propagating light can be a game changer for realizing a practical quantum computer. However, propagating light also has a drawback: the difficulty of logical encoding due to weak nonlinearity. Here, we propose Gaussian breeding that encodes arbitrary Gottesman-Kitaev-Preskill (GKP) qubits in propagating light. The key idea is the efficient and iterable generation of quantum superpositions by photon detectors, which is the most widely used nonlinear element in quantum propagating light. This formulation makes it possible to systematically create the desired qubits with minimal resources. Our simulations show that GKP qubits above a fault-tolerant threshold, including ``magic states'', can be generated with a high success probability and with a high fidelity exceeding 0.99. This result fills an important missing piece toward practical quantum computing.}}




\maketitle
Quantum computers are expected to outperform classical computers in certain tasks. For quantum computers to become a technology that changes our lives, it is necessary to protect fragile quantum information and ensure the reliability of computation. The basic idea for this is to encode quantum information redundantly as a logical qubit in a high-dimensional Hilbert space \cite{Gottesman2009, Gottesman1999}. However, encoding of logical qubits is typically challenging for any physical system, for different reasons. For example, scalability is a critical issue in stationary two-level systems, because each logical qubit should be encoded in a quantum many-body system. A logical qubit with high redundancy is encoded in $10^3$ to $10^4$ physical qubits, and millions or more physical qubits are required to perform practical tasks \cite{Fowler2012}. Spatial parallelization and control of such a large number of physical qubits are far beyond the current techniques that deal with tens to hundreds of physical qubits \cite{Preskill2018,Arute2019}.

For avoiding the problem of scalability, propagating light is a promising system. First, in principle, we no longer need a quantum many-body system for redundant encoding. Each logical qubit can be encoded in just one wave packet thanks to the infinite-dimensional Hilbert space of an oscillator \cite{Gottesman2001,Leghtas2013,Mirrahimi2014,Michael2016}. This advantage is common to any oscillator, but only propagating light has realized a scalable platform for large-scale quantum computing \cite{LarsenMikkel2019,Warit2019,Larsen2021}. This is because the flying qubits do not need to be spatially parallelized: they can be scalably multiplexed in the time domain. Despite this remarkable scalability, unavailable to other systems, the lack or natural weakness of nonlinear interactions with propagating light is limiting any logical encoding. It is known that nonlinear interactions between matter and light \cite{Travaglione2002,Fluehmann2019,Campagne-Ibarcq2020} or between light and light \cite{Gottesman2001,Fukui2022a} enable logical encoding, but these are challenging to employ in propagating light. Thus, leveraging of the nonlinearity of photon detectors, which has been widely used experimentally \cite{Harder2016,Becerra2015,Zhong2022,Madsen2022}, is a promising approach. However, it is highly nontrivial how to exploit the nonlinearity of photon detectors for logical encoding. 

Here, we propose a protocol for encoding arbitrary Gottesman-Kitaev-Preskill (GKP) qubits \cite{Gottesman2001} in propagating light using photon detectors. The GKP qubit is a powerful logical qubit, due to its intrinsic robustness, living in a quantum error correction code space. Ideally, the GKP qubit is a superposition of equally-spaced position eigenstates, but is usually approximated as a superposition of squeezed states, as shown in Fig. \ref{Fig:fig1}a. We shall analytically derive an operation that bifurcates a wave function by using a photon detector. By iterating the operation, we can realize the periodically discretized wave function of arbitrary GKP qubits, including so-called magic states as needed for universal quantum computing \cite{Bravyi2005,Knill2005}, with high fidelity and high success probability. Our novel formulation allows to incorporate the benefits of two distinct state-generation methods while circumventing their drawbacks: it is systematic and comprehensible like the breeding protocol \cite{Vasconcelos2010,Weigand2018} and has a feasible configuration like the Gaussian Boson sampling as a state synthesizer \cite{Su2019,Sabapathy2019a,Tzitrin2020,Fukui2022}. For this reason, we refer to the proposed protocol as Gaussian breeding.

\subsection*{Bifurcation of quantum states}
We introduce the basic operations of our protocol. We suppose that the position and momentum operators satisfy a commutation relation $[\hat{x},\hat{p}]=i$. For simplicity, normalization is omitted in the notation of quantum states below. We denote a squeezed vacuum state by $\ket{S_{\Delta}} = \hat{S}(\Delta)\ket{0}=e^{-i\frac{\ln{\Delta}}{2}(\hat{x}\hat{p}+\hat{p}\hat{x})}\ket{0}$, where the squeezing operator gives $\hat{S}^{\dag}(\Delta) \hat{x}\hat{S}(\Delta)= \hat{x}/\Delta$ and $\ket{0}$ is the single-mode vacuum state. The physical codewords of GKP qubits $\ket{\tilde{k}_{\Delta , \kappa}} \ (k=0,1)$ are
\begin{equation}\label{eq:GKP01}
\ket{\tilde{k}_{\Delta , \kappa}} = \Biggl[\sum_{s=-\infty}^{\infty} e^{-\frac{1}{2\kappa^2}\bigl((2s+k)\sqrt{\pi}\bigr)^2} \hat{D}\Bigl((2s+k)\sqrt{\pi}\Bigr)\Biggr] \ket{S_{\Delta}},
\end{equation}
where $\hat{D}(d)=e^{-id{\hat{p}}}\ (d\in \mathbb{R})$ is a position shift operator. These codewords approach the ideal GKP codewords in the limit of $\Delta, \kappa \rightarrow \infty$. We define {\it coherent bifurcation} $\mathcal{B}_{w}[\cdot]\ (w\in \mathbb{R})$ that satisfies
\begin{gather}
\mathcal{B}_{w}\left[ \hat{D}(d)\ket{S_{\Delta}} \right] = \left[ \hat{D}(d-w) + \hat{D}(d+w)  \right] \ket{S_{\Delta}},\label{eq:bfc_state} \\
\mathcal{B}_{w}\Bigl[ a\ket{\psi}+b\ket{\phi} \Bigr] = a\mathcal{B}_{w}\Bigl[ \ket{\psi} \Bigr] + b\mathcal{B}_{w}\Bigl[\ket{\phi} \Bigr]. \label{eq:linearity}
\end{gather}
We denote an $N$-times iteration of coherent bifurcation as $\mathcal{B}_{w}^{(N)}$,
\begin{gather}
\mathcal{B}_{w}^{(N)}\left[ \ket{\psi} \right] = \mathcal{B}_{w}\Bigl[ \mathcal{B}_{w}^{(N-1)}\left[ \ket{\psi} \right] \Bigr],\ \mathcal{B}_{w}^{(0)}\left[ \ket{\psi} \right] = \ket{\psi}.
\end{gather}
As shown in Fig. \ref{Fig:fig1}b and Methods, the GKP codewords can be obtained from a squeezed vacuum state via suitable $N$-times coherent bifurcation,
\begin{equation}\label{eq:gen_of_codewords}
\mathcal{B}_{\sqrt{\pi}}^{(N)}\Bigl[ \ket{S_{\Delta}} \Bigr] 
\approx \ket{\tilde{k}_{\Delta , \sqrt{N\pi}}},\ k \equiv N \ ({\rm mod}\ 2).
\end{equation}
Arbitrary GKP qubits can be generated from a seed state defined by
\begin{equation}
\ket{{\rm seed}_{\Delta, \alpha, \beta}} = \alpha\ket{S_{\Delta}}+\frac{\beta}{2}\ \cdot\mathcal{B}_{\sqrt{\pi}}\Bigl[ \ket{S_{\Delta}} \Bigr],
\end{equation}
where $\abs{\alpha}^2+\abs{\beta}^2=1$ and $\abs{\alpha}\ge \abs{\beta}$. By applying an appropriate $\mathcal{B}_{w}^{(N)}$, now we get
\begin{equation}\label{Eq:arbitrary_GKP}
\mathcal{B}_{\sqrt{\pi}}^{(N)}\Bigl[ \ket{{\rm seed}_{\Delta, \alpha, \beta}} \Bigr]
\approx \hat{X}^N \Bigl[ \alpha\ket{\tilde{0}_{\Delta, \sqrt{N\pi}}}+\beta\ket{\tilde{1}_{\Delta, \sqrt{N\pi}}} \Bigr],
\end{equation}
as shown in Fig. \ref{Fig:fig1}c and Methods, where $\hat{X}$ is a logical bit-flip operator. One possibility to provide the seed states is to make use of a damping operation $e^{-t\hat{x}^2}$ ($e^{-t\hat{p}^2}$), which multiplies $e^{-t x^2}$ ($e^{-t p^2}$) to the wave function of $x$ ($p$). As shown in Fig. \ref{Fig:fig1}d, seed states can thus be generated as
\begin{equation}
\ket{{\rm seed}_{\Delta, \alpha, \beta}} \approx \hat{S}(\delta)\cdot e^{-c{\hat{p}}^2}\cdot e^{ib \hat{x}^2}\cdot e^{-a{\hat{x}}^2}\cdot \mathcal{B}_{w}^{(2)}\Bigl[\ket{S_{\Delta}}\Bigr]\ \ \ \ (w\ge \sqrt{\pi}/2), \label{eq:seed_generation}
\end{equation}
with certain $a,b,c,$ and $\delta$ given in Methods. The approximation gets better as $w$ increases. We can therefore generate the arbitrary GKP qubits using the two nontrivial operations: coherent bifurcation and damping.

\subsection*{Implementation with minimal resources}
The above scheme resembles the breeding protocol \cite{Vasconcelos2010,Weigand2018} in that it systematically generates the desired states by iterating elementary operations. However, we note that the conventional breeding is demanding in terms of resource requirements. It can generate the codewords of GKP qubits by iterating the interference of Schr\"{o}dinger cat states followed by an amplitude measurement. The interval of the superposed squeezed states gets smaller after each breeding step, so we need to increase the amplitude of the cat states exponentially with the iteration number. In addition, performing amplitude measurements on non-Gaussian states is inefficient, because the usual breeding scheme also involves photon number measurements for the generation of the initial non-Gaussian states.

Reflecting on this, we propose a minimal configuration of our breeding protocol, which only performs photon number measurements on a Gaussian state. This configuration can be analogized to the Gaussian Boson Sampling \cite{Hamilton2017}, so we refer to our protocol as Gaussian breeding. To understand how to realize it, we decompose Eq. (\ref{eq:bfc_state}) into two equations,
\begin{gather}
\mathcal{B}_{w}\left[ \ket{S_{\Delta}} \right] = \left[ \hat{D}(-w) + \hat{D}(w)  \right] \ket{S_{\Delta}}, \label{eq:def_decomp_1} \\
\mathcal{B}_{w}\left[\hat{D}(d)\ket{S_{\Delta}} \right]=\hat{D}(d)\mathcal{B}_{w}\Bigl[\ket{S_{\Delta}} \Bigr]. \label{eq:def_decomp_2}
\end{gather}
Equation (\ref{eq:def_decomp_1}) is realized by a generalized photon subtraction \cite{Takase2021}, a method for creating Schr\"{o}dinger cat states. Figure \ref{Fig:fig1}e shows the setup, where two squeezed vacuum states $\ket{S_{\Delta_1}}_1$ and $\ket{S_{\Delta_2}}_2$ interact by a beam splitter interaction $\hat{B}$ with transmittance $T=1-R$ followed by the detection of $n$ photons. Any $n$ is useful, but below we assume $n$ is even. This process $\mathcal{G}_w$ well approximates Eq. (\ref{eq:def_decomp_1}),
\begin{equation}
\mathcal{G}_{w}\Bigl[\ket{S_{\Delta}} \Bigr] \approx \left[ \hat{D}(-w) + \hat{D}(w)  \right] \ket{S_{\Delta}},\ w=\sqrt{2n}\Delta^{-1}, \label{eq:B_G}
\end{equation}
when $n\ge 2$, $T\Delta_1^2 + R\Delta_2^2 =1$ and $\Delta_2^2=1/(1+\Delta_1^2)$ (see Methods). Compared to conventional photon subtraction \cite{Dakna1997}, the generalized photon subtraction can achieve a larger $w$ for the same $n$ with a much higher success probability, which indicates that the nonlinearity of the photon detector is efficiently exploited. Unfortunately, though, $\mathcal{G}_w$ does not in general implement $\mathcal{B}_w$, because the non-commutativity $\left[\hat{D}_1(d),\hat{B} \right]\neq 0$ leads to
\begin{equation}
\mathcal{G}_{w}\left[\hat{D}(d)\ket{S_{\Delta}} \right]\neq \hat{D}(d)\mathcal{G}_{w}\Bigl[\ket{S_{\Delta}} \Bigr],
\end{equation}
which contradicts Eq. (\ref{eq:def_decomp_2}). We can avoid this problem by using a different interaction. We propose the iterable generalized photon subtraction $\mathcal{\tilde{G}}_{w}$ in Fig. \ref{Fig:fig1}f, where the beam splitter is replaced by a quantum-non-demolition interaction $\hat{Q}(g)=e^{ig\hat{p}_1\hat{x}_2}$ and $\hat{S}_2(\Delta_3)$. Since $\left[\hat{S}_2(\Delta_3)\hat{Q}(g), \hat{D}_1(d) \right]= 0$, now we do get
\begin{equation}\label{eq:g_tilde}
\mathcal{\tilde{G}}_{w}\left[\hat{D}(d)\ket{S_{\Delta}} \right] = \hat{D}(d)\mathcal{\tilde{G}}_{w}\Bigl[\ket{S_{\Delta}} \Bigr],
\end{equation}
as required according to Eq. (\ref{eq:def_decomp_2}). The operation $\mathcal{\tilde{G}}_{w}$ equivalently approximates Eq. (\ref{eq:def_decomp_1}) as $\mathcal{G}_w$ when $n\ge 2$, $\left(\Delta_2^{2}+g^2\Delta_1^2\right)\Delta_3^2 =1$ and $\Delta_2^2\ll 1 \ll \Delta_1^2$ (see Methods). This process also obeys the linearity as shown in Eq. (\ref{eq:linearity}). As a result, $\tilde{\mathcal{G}}_w$ is an implementation of coherent bifurcation $\mathcal{B}_w$. This operation inherits the advantages of the generalized photon subtraction, such as large $w$, a high success probability, and a high fidelity. Interestingly, this setup also realizes the damping operation $e^{-t\hat{p}^2}$ when $n=0$. We can further realize a damping about $x$ by applying a $\pi/2$ phase rotation of the input and output states. Therefore, we are able to generate arbitrary GKP qubits by cascading the optical circuit in Fig. \ref{Fig:fig1}f.

For implementing the Gaussian breeding, plenty of on-line squeezers as needed in the Gaussian operations seem to pose an experimental challenge \cite{Filip2005,Shiozawa2018}. Using Bloch-Messiah reduction \cite{Braunstein2005}, however, we can find an equivalent circuit that only uses off-line squeezers, beam splitters, and photon number measurements. This configuration, as shown in Fig. \ref{Fig:fig1}g, automatically determining the specific parameters of Gaussian Boson sampling as a state synthesizer, is much more preferable for experimental feasibility \cite{Zhong2022,Madsen2022}. Note that Gaussian Boson sampling has a complexity enough to achieve quantum supremacy \cite{Hamilton2017,Zhong2022,Madsen2022}. It is therefore difficult to find practical parameters in a realistic time. The best we could do with the original type of state synthesizer is to numerically obtain the parameters that generate the codewords with a limited fidelity \cite{Tzitrin2020} or with a very low probability \cite{Fukui2022}. The Gaussian breeding systematically and analytically solves this problem by introducing a decomposition of GKP-qubit generation based on the concept of coherent bifurcation. The number of beam splitters is at most $N(N+1)/2$ when we perform photon number measurements $N$ times \cite{Reck1994}. In practice, this scaling can be improved by choosing optical parameters. For example, when we generate the codeword states by iterating completely the same coherent bifurcation $N$ times, $N$ beam splitters are actually sufficient \cite{Loock2007} as shown in Fig. \ref{Fig:fig1}h.

\subsection*{Performance}
We conduct numerical simulations of GKP-qubit generation in a Fock space up to 55 photons. First, we simulate generation of the codewords $\ket{\tilde{0}_{\Delta , \Delta}}, \ket{\tilde{1}_{\Delta , \Delta}}$. We consider three cases where coherent bifurcation with $n=6,10,16$ is iteratively employed. We perform a damping operation after iterating the bifurcation $N$ times to increase the fidelity to the target states. Table \ref{tab0} shows the results of the simulation. We can see that a larger $\Delta$ is achieved by increasing the value of $n$. When $n=16$, $\Delta$ exceeds $\sqrt{10}$ (10 dB squeezing), which is usually  assumed as the fault-tolerant threshold \cite{Fukui2018}. All the simulated states have a high fidelity $F>0.99$. Conventional breeding can generate similar states, but each breeding step reduces the interval between squeezed states by $1/\sqrt{2}$. As a result, the number of detected photons increases exponentially in conventional breeding with the number of iterations, whereas it increases only linearly in the Gaussian breeding. For example, to achieve $\Delta \ge \sqrt{10}$ requires detection of 256 photons with conventional breeding, 4 times larger than for the Gaussian breeding. In a numerical analysis of state synthesizers based on concepts of similar to Gaussian Boson sampling, $\Delta \ge \sqrt{10}$ and a high fidelity $F>0.999$ are achieved, but the success probability is kept as low as $10^{-29}$ \cite{Fukui2022}. These comparisons indicate the high efficiency of the Gaussian breeding. Figures \ref{Fig:fig2}a-c show the Wigner functions of the simulated $\ket{\tilde{0}_{\Delta , \Delta}}$ states. These Wigner functions illustrate well the characteristics of GKP qubits that are often referred to as grid states. We can see that the gird structure becomes clearer as $n$ increases.

Next, we simulate generation of arbitrary GKP qubits $\alpha \ket{\tilde{0}_{\Delta , \Delta}} + \beta e^{i\phi} \ket{\tilde{1}_{\Delta , \Delta}}$. We target the three magic states \cite{Gottesman2001,Bravyi2005,Knill2005} with $\left(\alpha,\beta \right)=\left(\cos{\frac{\pi}{8}},\sin{\frac{\pi}{8}}\right),\left( \frac{1}{\sqrt{2}},\frac{e^{-i\frac{\pi}{4}}}{\sqrt{2}} \right)$, and $\left( \cos{\theta},\sin{\theta}e^{-i\frac{\pi}{4}} \right)$ where $\cos{2\theta}=1/\sqrt{3}$. Table \ref{tab0} shows the simulation results. Magic states exceeding the fault-tolerant threshold are generated with a high fidelity $F>0.99$. The success probability is smaller than for the case of codeword generation, because now we have to include the probability of generating seed states, which is about $10^{-5}$. The Wigner functions of the seed states are shown in Figs. \ref{Fig:fig2}d-f. Figures \ref{Fig:fig2}g-i present the Wigner functions of the simulated magic states. We can see that more complicated grid structures compared to the standard codewords can be successfully synthesized. The significance of this is that, in particular, resources of GKP magic states can be exploited to do universal quantum computation \cite{Baragiola2019,Yamasaki2020} by means of non-Clifford gate teleportation \cite{Konno2021}.

\subsection*{Outlook}
Since a specific recipe for GKP-qubit generation has been presented, an experimental verification could be actively pursued. For the time being, GKP qubits with small $\Delta$ and $\kappa$ would be generated by detecting a small number of photons. Injecting these states into quantum processors \cite{Takase2022a} and using them for a proof-of-principle demonstration of quantum error correction \cite{Walshe2020} and non-Clifford operations \cite{Konno2021,Sakaguchi2022} are an important research direction. Generating more advanced states would require exploiting the broadband nature of light. In principle, the attempt rate of state generation can be enhanced as high as the bandwidth of the input squeezed vacuum states, which can be 10 THz with current technology \cite{Kashiwazaki2021b}. Thus, a success probability of $10^{-10}$ is high enough for demonstrating state generation, and even quasi-deterministic generation is possible by using a quantum memory with a reasonable life time of $1$ ms \cite{Rancic2018}. Other factors such as dead time of photon detectors limit the attempt rate, but even in that case wavelength-division-multiplexed state generation \cite{Joshi2018} could exploit the full bandwidth of the squeezed light. Although the original proposal utilizes on-line nonlinear elements for wavelength conversion, quantum teleportation across different wavelength bands would be another feasible option.

There are also expectations for theoretical improvements. The restrictions $\left(\Delta_2^{2}+g^2\Delta_1^2\right)\Delta_3^2 =1, \Delta_2^2\ll 1 \ll \Delta_1^2$, and detection of the same $n$ in each coherent bifurcation make our method simple, but are not always necessary. By relaxing these restrictions, we would achieve a higher success probability and a higher fidelity with detection of fewer photons. The Gaussian breeding, which combines small functional blocks, is expected to be highly compatible with a parameter-search method called back casting \cite{Fukui2022}. Incorporating such an analysis method is one promising direction. Improving the fault-tolerant threshold also has a significant impact on feasibility. There is a proposal for a threshold of 8.3 dB \cite{Fukui2019arxiv} instead of 10 dB, which reduces the number of detected photons required for each coherent bifurcation from $n=16$ to $n=10$. The Gaussian breeding, as proposed in this work, merging the systematic iterative circuitry of conventional breeding with the experimental feasibility of Gaussian Boson sampling, will become a more powerful method when combined with the progress of related research as mentioned above, and will become a driving force for the realization of a practical optical quantum computer.

\section*{Methods}

\subsection*{Coherent bifurcation}
Here, we derive Eqs. (\ref{eq:gen_of_codewords}) and (\ref{Eq:arbitrary_GKP}). From the properties of $\mathcal{B}_{w}^{(N)}$, we get
\begin{equation}\label{eq:bfc_binomial_codewords}
\mathcal{B}_{w}^{(N)}\Bigl[ \ket{S_{\Delta}} \Bigr] = \left[ \sum_{l=0}^{N} \ _{N}C_l \cdot \hat{D}\Bigl((2l-N)w\Bigr) \right] \ket{S_{\Delta}}.
\end{equation}
When $N\gg 1$, the binomial coefficient $_{N}C_l$ is well approximated by a Gaussian function,
\begin{equation}
_{N}C_l \approx \sqrt{\frac{2^{2N+1}}{N\pi}}\exp{\left(-\frac{2(l-N/2)^2}{N}\right)}.
\end{equation}
When each displaced squeezed state in Eq. (\ref{eq:bfc_binomial_codewords}) is enough separated, we have
\begin{equation}
\abs{\mathcal{B}_{w}^{(N)}\Bigl[ \ket{S_{\Delta}} \Bigr]}^2 = \sum_{l=0}^{N} \left(_NC_l \right)^2= {}_{2N}C_N \approx \frac{2^{2N}}{\sqrt{\pi}}.
\end{equation}
When $N\gg 1$ and $w=\sqrt{\pi}$, we get
\begin{align}\label{eq:codewords_detail}
\frac{\sqrt[4]{\pi}}{2^{N}}\ \mathcal{B}_{\sqrt{\pi}}^{(N)}\Bigl[ \ket{S_{\Delta}} \Bigr] &\approx \sqrt{\frac{2}{N\sqrt{\pi}}}\left[ \sum_{l=0}^{N} \exp{\left(-\frac{\Bigl((2l-N)\sqrt{\pi}\Bigr)^2}{2N\pi}\right)}  \cdot \hat{D}\Bigl((2l-N)\sqrt{\pi}\Bigr) \right] \ket{S_{\Delta}} \nonumber \\
&\approx \ket{\tilde{k}_{\Delta , \sqrt{N\pi}}},\ k \equiv N \ ({\rm mod}\ 2).
\end{align}
The generation of arbitrary GKP qubits is confirmed from Eq. (\ref{eq:codewords_detail}), 
\begin{align}
\sqrt{\frac{\sqrt{\pi}}{2^{2N}}}\ \mathcal{B}_{\sqrt{\pi}}^{(N)}\Bigl[ \ket{{\rm seed}} \Bigr]
&= \alpha \sqrt{\frac{\sqrt{\pi}}{2^{2N}}}\ \mathcal{B}_{\sqrt{\pi}}^{(N)}\Bigl[ \ket{S_{\Delta}} \Bigr]+ \beta \sqrt{\frac{\sqrt{\pi}}{2^{2(N+1)}}}\ \mathcal{B}_{\sqrt{\pi}}^{(N+1)}\Bigl[ \ket{S_{\Delta}} \Bigr] \nonumber \\
&\approx
\hat{X}^N \Bigl[ \alpha\ket{\tilde{0}_{\Delta, \sqrt{N\pi}}}+\beta\ket{\tilde{1}_{\Delta, \sqrt{N\pi}}} \Bigr].
\end{align}

\subsection*{Damping operation}
The damping operation is realized by supposing $n=0$ in Fig. \ref{Fig:fig1}f, where $\hat{Q}(g)=e^{ig\hat{p}_1\hat{x}_2}$. When the input state is $\ket{\psi_{{\rm in}}}$, the output state is
\begin{equation}
\ket{\Psi_{{\rm out}}}_1 = _2\bra{0}\hat{S}_2(\Delta_3)\hat{Q}(g) \ket{\psi_{{\rm in}}}_1 \ket{S_{\Delta_2}}_2.
\end{equation}
Note that $\ket{\Psi_{{\rm out}}}_1$ is not normalized. The wave function of this state is
\begin{align}
\Psi_{{\rm out}}(p_1) &= \int dp_2\ \psi_{{\rm in}}(p_1)e^{-\frac{1}{2\Delta_2^2\Delta_3^2}(p_2 - \Delta_3 gp_1)^2}\cdot e^{-\frac{1}{2}p_2^2} \nonumber \\
&= \sqrt{\frac{2\Delta_2^2\Delta_3^2}{1+\Delta_2^2\Delta_3^2}\pi}\ \psi_{{\rm in}}(p_1)\cdot \exp{\left[ -\frac{1}{2}\cdot \frac{g^2\Delta_3^2}{1+\Delta_2^2\Delta_3^2}p_1^2 \right]}.
\end{align}
This process therefore implements a damping operation $e^{-t\hat{p}^2}$ with $t = \frac{g^2\Delta_3^2}{2(1+\Delta_2^2\Delta_3^2)}$. We can also realize a damping about $x$ by applying a $\pi/2$ phase rotation of the input and output states.

The parameters in Eq. (\ref{eq:seed_generation}) explicitly are
\begin{equation}
a = \frac{1}{4w^2}\ln{\abs{\frac{\alpha}{\beta}}},\ b = \frac{1}{4w^2}\arg{\left( \frac{\beta}{\alpha} \right)},\ c = \frac{1}{2\Delta^2}\left( \frac{4w^2}{\pi}-1 \right),\ \delta = \frac{2w}{\sqrt{\pi}}.
\end{equation}
In Eq. (\ref{eq:seed_generation}), we can omit the term $\hat{S}(\delta)\cdot e^{-c{\hat{p}}^2}$ when $w=\sqrt{\pi}/2$, but more accurate seed states are obtained as $w$ becomes larger than $\sqrt{\pi}/2$. This is because the term $e^{ib \hat{x}^2}\cdot e^{-a{\hat{x}}^2}$, which is intended to multiply $\beta/\alpha$ to the displaced terms of $\mathcal{B}_{w}^{(2)}\Bigl[\ket{S_{\Delta}}\Bigr]=\Bigl[\hat{D}(-2w) + 2 + \hat{D}(2w) \Bigr]\ket{S_{\Delta}}$, works more accurately. Instead, we have to adjust the variance and displacement of the squeezed states when $w>\sqrt{\pi}/2$. The damping $e^{-c{\hat{p}}^2}$ increases the variance of the squeezed states by convolving a Gaussian $e^{-\frac{1}{4c}x^2}$ to the wave function about $x$. Finally, we get the desired state by applying $\hat{S}(\delta)$. Besides the generation of seed states, the damping operation also can be used to adjust the envelope of the GKP qubits. When $\sqrt{N\pi} > \Delta$, we can get square-lattice GKP qubits, $e^{-t\hat{x}^2}\left[\alpha \ket{\tilde{0}_{\Delta , \sqrt{N}w}} + \beta \ket{\tilde{1}_{\Delta , \sqrt{N}w}}\right] \approx \alpha \ket{\tilde{0}_{\Delta , \Delta}} + \beta \ket{\tilde{1}_{\Delta , \Delta}}$, up to normalization by choosing a proper $t$.

\subsection*{Iterable generalized photon subtraction}
\begin{figure*}[t]
	\begin{center}
		\includegraphics[bb= 0 0 510 610,clip,width=\textwidth]{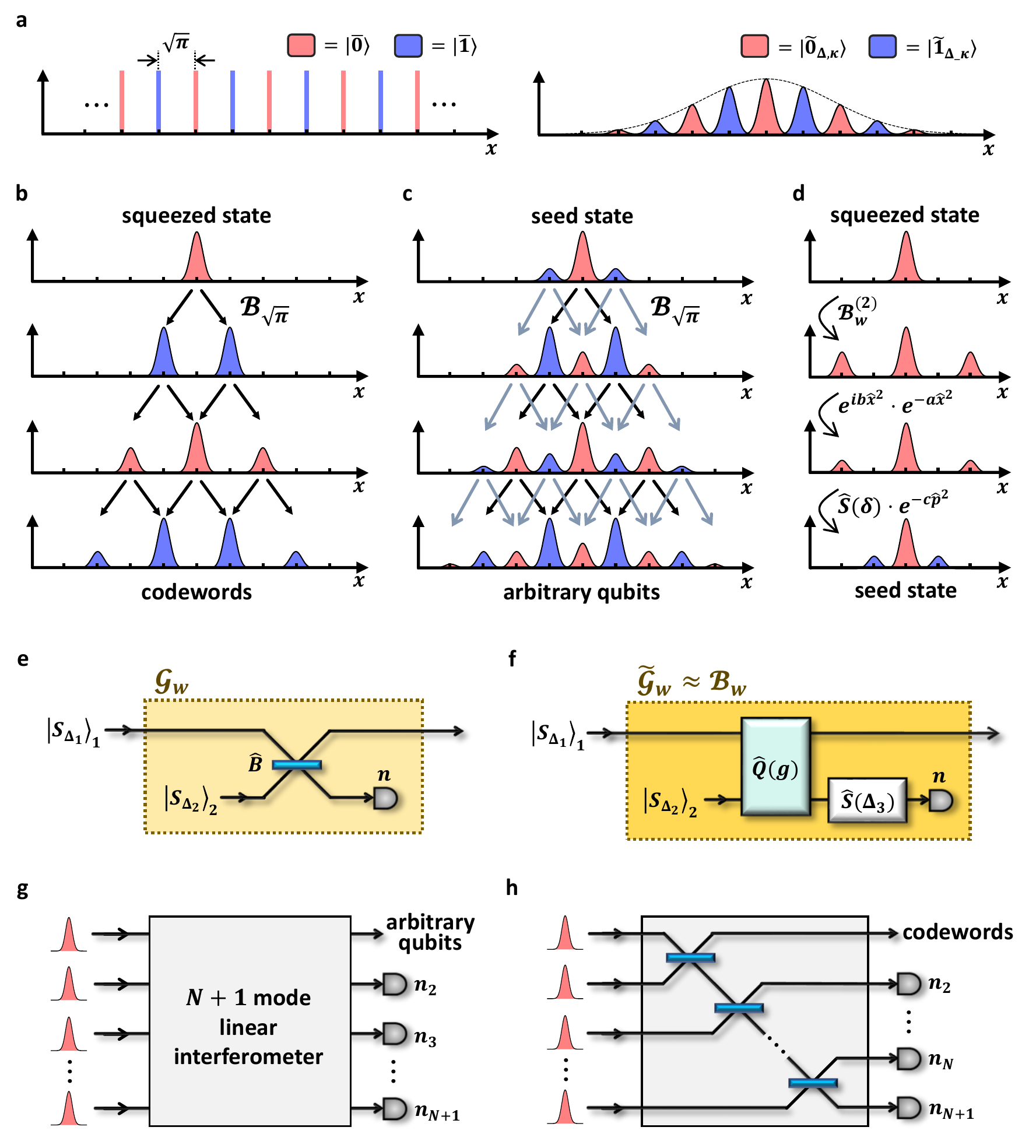}
	\end{center}
	\caption{{\bf Schematic of the Gaussian breeding. (a)} Wave functions of GKP qubits. Ideal codewords $\ket{\bar{0}}, \ket{\bar{1}}$ are a superposition of position eigenstates spaced by $2\sqrt{\pi}$. The approximated codewords $\ket{\tilde{0}_{\Delta,\kappa}}, \ket{\tilde{1}_{\Delta,\kappa}}$ are a superposition of squeezed states (variance $1/2\Delta^2$) with a Gaussian envelope (variance $\kappa^2/2$). {\bf (b)} Generation of the codewords by iterative coherent bifurcation on a squeezed state. {\bf (c)} Generation of arbitrary GKP qubits by using the seed state as an initial state. {\bf (d)} Generation of a seed state from a squeezed state using coherent bifurcation, damping, and Gaussian operations. This diagram shows the case of $w=\sqrt{\pi}$. {\bf (e)} A setup of the generalized photon subtraction. Squeezed Schr\"{o}dinger cat states are generated from squeezed states. {\bf (f)} The setup of the iterable generalized photon subtraction using a quantum-non-demolition interaction. It implements the coherent bifurcation and the damping operation only with the help of a Gaussian ancillary state, Gaussian operations, and a photon number measurement. {\bf (g)} Gaussian breeding in a form of Gaussian Boson sampling. The Gaussian breeding can generate arbitrary GKP qubits by cascading the optical circuit in (f) and a few Gaussian operations. In experiments, the configuration shown in this figure, which is derived from the Bloch-Messiah reduction, is more advantageous. {\bf (h)} Generation of the codewords in a state-synthesizer configuration. $N$ beam splitters are enough for an $N$-times iteration of the coherent bifurcation.}
\label{Fig:fig1}
\end{figure*}

\begin{figure*}[t]
	\begin{center}
		\includegraphics[bb= 0 0 510 470,clip,width=\textwidth]{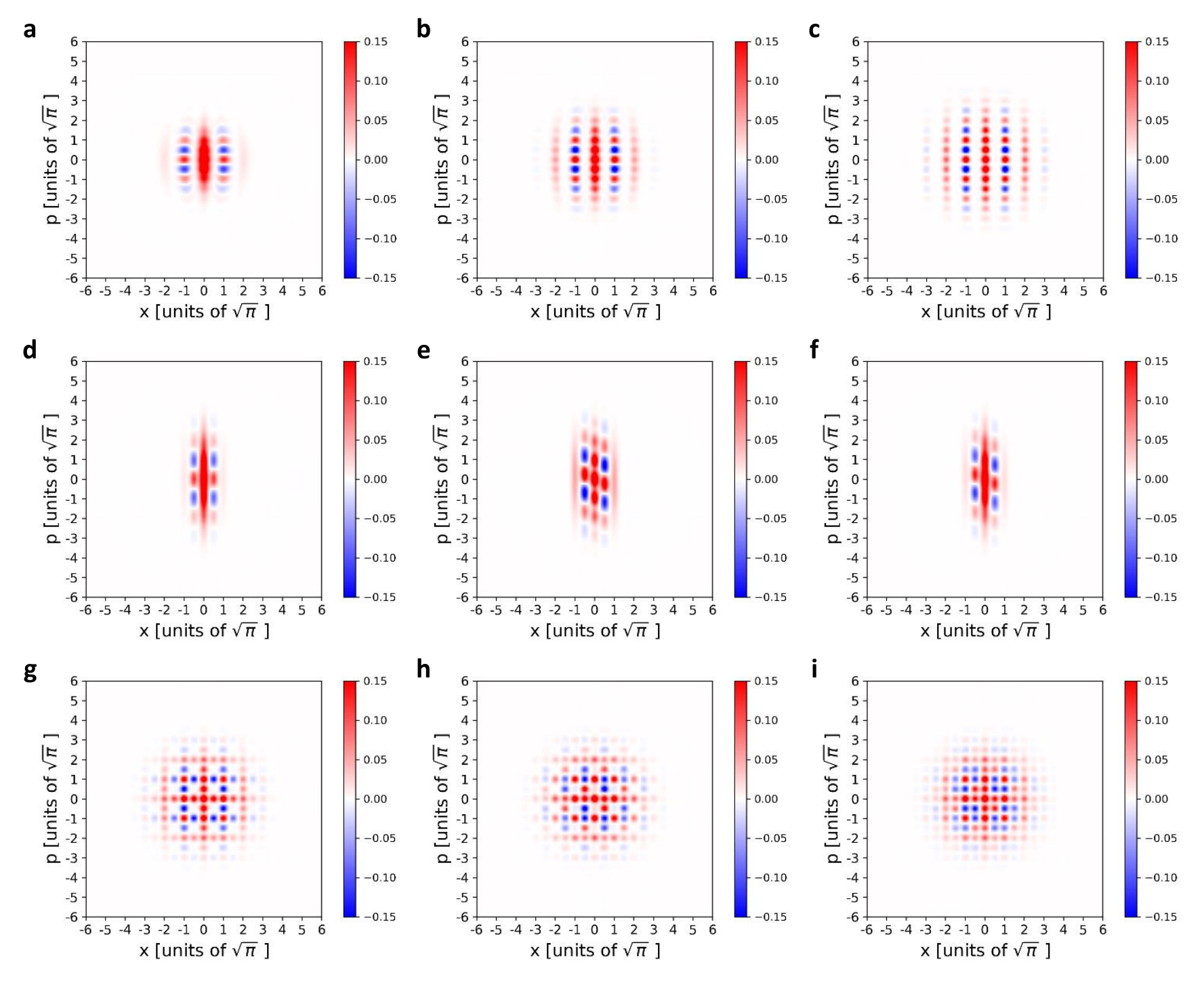}
	\end{center}
	\caption{{\bf Wigner functions of the simulated states. (a)-(c)} Results of the codeword generation. Each state corresponds to the condition of $(n, N)=(6,2), (10,4), (16, 4)$. {\bf (d)-(f)} Results of the seed-state generation. Each state corresponds to $\left(\alpha,\beta \right)=\left(\cos{\frac{\pi}{8}},\sin{\frac{\pi}{8}}\right),\left( \frac{1}{\sqrt{2}},\frac{e^{-i\frac{\pi}{4}}}{\sqrt{2}} \right)$, and $\left( \cos{\theta},\sin{\theta}e^{-i\frac{\pi}{4}} \right)$ where $\cos{2\theta}=1/\sqrt{3}$. {\bf (g)-(i)} Results of the seed-state generation. Each state is generated from the seed state shown in (d)-(f), respectively. Coherent bifurcation is performed on the seed states 4 times with $n=16$.}
\label{Fig:fig2}
\end{figure*}

The generalized photon subtraction \cite{Takase2021} is a protocol to generate Schr\"{o}dinger cat states. Figure \ref{Fig:fig1}e is the original setup defining the single-mode operation $\mathcal{G}_{w}$. We propose a new setup in Fig. \ref{Fig:fig1}f as a different implementation defining the modified, adapted single-mode operation $\tilde{\mathcal{G}}_{w}$, which can be used for the Gaussian breeding. As the basic properties common to the both cases, the output state is given by
\begin{equation}
\ket{\Psi_n}_1 = \ _2\bra{n}\ket{G}_{1,2},
\end{equation}
where $\ket{G}_{1,2}$ is a two-mode Gaussian state. Note that $\ket{\Psi_n}_1$ is not normalized. The wave function $G(x_1,x_2) =\ _2\bra{x_2}\ _1\bra{x_1}\ket{G}_{1,2}$ can be an arbitrary Gaussian function but it is enough to assume the following form,
\begin{equation}
G(x_1,x_2)=\frac{\abs{\sigma}^{\frac{1}{4}}}{\sqrt{\pi}} \exp{\left[ -\frac{1}{2}
\bm{x}^T
\sigma
\bm{x}
 \right]},\ 
\bm{x} =
\begin{pmatrix}
x_1 \\
x_2 \\
\end{pmatrix},
\end{equation}
where $\sigma$ is a $2\times 2$ real symmetric matrix. Assuming $\sigma_{22}=1$ makes it easier to handle this protocol analytically. First, we get
\begin{gather}
\Psi_n(x_1) = \left( \frac{\abs{\sigma}}{\pi} \right)^{\frac{1}{4}}\frac{\left(-\sigma_{12}\right)^n}{\sqrt{2^nn!}}\ x_1^n\ {\rm exp}\left( -\frac{\Delta_c^2}{4}x_1^2 \right), \label{xexp} \\
\Delta_c^2\ =\ \abs{\sigma}+\sigma_{11}. \label{delta_c}
\end{gather}
The function in Eq. (\ref{xexp}) is well approximated by superimposed Gaussian functions,
\begin{equation}\label{eq:wf_approx}
\mathcal{N}\left[\Psi_n(x_1)\right] \approx \mathcal{N}\left[ {\rm exp}\left(- \frac{\Delta_c^2}{2}\left(x_1-\frac{\sqrt{2n}}{\Delta_c}\right)^2 \right) +(-1)^n  {\rm exp}\left(- \frac{\Delta_c^2}{2}\left(x_1+\frac{\sqrt{2n}}{\Delta_c}\right)^2 \right) \right],
\end{equation}
where $\mathcal{N}[\cdot]$ represents normalization. The fidelity of this approximation is
\begin{equation}
F_n = \frac{2^{n+\frac{5}{2}}e^{-\frac{2n}{3}}n!\abs{H_n\left(i\sqrt{\frac{2n}{3}}\right)}}{3^{n+1}(2n)!\left[1+(-1)^ne^{-2n}\right]} \approx 1-0.03/n.
\end{equation}
The probability to detect $n$ photons is
\begin{equation}
P(n) = \int \abs{\Psi_n(x)}^2 \ dx = \frac{\sqrt{2}(2n)!}{4^n (n!)^2}t^n(t+2)^{-n-\frac{1}{2}},\ t=g^2\frac{\Delta_1}{\Delta_2}.
\end{equation}
When $t=4n$, $P(n)$ has a maximum value given by
\begin{equation}
P_{{\rm max}}(n) = \frac{(2n)!}{2^n (n!)^2}\sqrt{\frac{1}{2n+1}}\left( \frac{n}{2n+1} \right)^n.
\end{equation}

In Figs. \ref{Fig:fig1}e,f, the squeezed state $\ket{S_{\Delta_1}}_1$ can be regarded as the input state. From Eq. (\ref{eq:wf_approx}), we can realize Eq. (\ref{eq:def_decomp_1}) when the following conditions are satisfied,
\begin{equation}
\sigma_{22}=1,\ \ \Delta_c^2\ =\ \abs{\sigma}+\sigma_{11}=\Delta_1^2. \label{eq:eq9cond}
\end{equation}
Let us derive a more specific expression for these conditions. The elements of $\sigma$ can be easily obtained from the calculation rule of covariance matrices. In the case of Fig. \ref{Fig:fig1}e, the state $\ket{G}_{1,2}=\hat{B} \ket{S_{\Delta_1}}_1 \ket{S_{\Delta_2}}_2$ gives
\begin{gather}
\sigma^{-1} =
\begin{pmatrix}
\sqrt{R} & -\sqrt{T} \\
\sqrt{T} & \sqrt{R} \\
\end{pmatrix}
\begin{pmatrix}
\Delta_1^{-2} & 0 \\
0 & \Delta_2^{-2} \\
\end{pmatrix}
\begin{pmatrix}
\sqrt{R} & \sqrt{T} \\
-\sqrt{T} & \sqrt{R} \\
\end{pmatrix}
=
\begin{pmatrix}
R\Delta_1^{-2}+T\Delta_2^{-2} & \sqrt{RT}\left( \Delta_1^{-2}-\Delta_2^{-2} \right) \\
\sqrt{RT}\left( \Delta_1^{-2}-\Delta_2^{-2} \right) & T\Delta_1^{-2}+R\Delta_2^{-2} \\
\end{pmatrix}, \\
\sigma =
\begin{pmatrix}
R\Delta_1^{2}+T\Delta_2^{2} & \sqrt{RT}\left( \Delta_1^{2}-\Delta_2^{2} \right) \\
\sqrt{RT}\left( \Delta_1^{2}-\Delta_2^{2} \right) & T\Delta_1^{2}+R\Delta_2^{2} \\
\end{pmatrix}.
\end{gather}
Thus, Eq. (\ref{eq:eq9cond}) is given by
\begin{equation}
T\Delta_1^{2}+R\Delta_2^{2}=1,\ \ \Delta_2^2 = \frac{1}{1+\Delta_1^2}.
\end{equation}
Under these conditions, the process in Fig. \ref{Fig:fig1}e is given by Eq. (\ref{eq:B_G}). Similarly, in the case of Fig. \ref{Fig:fig1}f, the state $\ket{G}_{1,2} = \hat{S}_2(\Delta_3)\hat{Q}(g) \ket{S_{\Delta_1}}_1 \ket{S_{\Delta_2}}_2$ gives
\begin{gather}
\sigma^{-1} =
\begin{pmatrix}
1 & 0 \\
0 & \Delta_3^{-1} \\
\end{pmatrix}
\begin{pmatrix}
1 & g \\
0 & 1 \\
\end{pmatrix}
\begin{pmatrix}
\Delta_1^{-2} & 0 \\
0 & \Delta_2^{-2} \\
\end{pmatrix}
\begin{pmatrix}
1 & 0 \\
g & 1 \\
\end{pmatrix}
\begin{pmatrix}
1 & 0 \\
0 & \Delta_3^{-1} \\
\end{pmatrix}
=
\begin{pmatrix}
\Delta_1^{-2}+g^2\Delta_2^{-2} & g\Delta_2^{-2}\Delta_3^{-1} \\
g\Delta_2^{-2}\Delta_3^{-1} & \Delta_2^{-2}\Delta_3^{-2} \\
\end{pmatrix}, \\
\sigma =
\begin{pmatrix}
\Delta_1^{2} & -g\Delta_1^2\Delta_3 \\
-g\Delta_1^2\Delta_3 & \left(\Delta_2^{2}+g^2\Delta_1^2\right)\Delta_3^2 \\
\end{pmatrix}.
\end{gather}
Thus, Eq. (\ref{eq:eq9cond}) is given by
\begin{equation}
\left(\Delta_2^{2}+g^2\Delta_1^2\right)\Delta_3^2=1,\ \ \Delta_1^2\Delta_2^2\Delta_3^2=0.
\end{equation}
Note that $\Delta_1^2\Delta_2^2\Delta_3^2=0$ is unphysical, thus this condition should be approximately satisfied by assuming $\Delta_2^2\ll 1 \ll \Delta_1^2$ in addition to the former condition. With these conditions, the process in Fig. \ref{Fig:fig1}f is given by
\begin{equation}
\mathcal{\tilde{G}}_w\Bigl[ \ket{S_{\Delta}} \Bigr] \approx \left[ \hat{D}(-w) + \hat{D}(w)  \right] \ket{S_{\Delta}},\ w=\sqrt{2n}\Delta^{-1}.
\end{equation}
From the above, both setups in Figs. \ref{Fig:fig1}e,f can realize Eq. (\ref{eq:def_decomp_1}). However, only Fig. \ref{Fig:fig1}f satisfies Eq. (\ref{eq:def_decomp_2}), as discussed in the main text.

\backmatter

\bmhead{Acknowledgments}
This work was partly supported by the Japan Society for the Promotion of Science KAKENHI (18H05207, 20K15187, and 22K20351), the Japan Science and Technology Agency (JPMJMS2064), the BMBF in Germany (QR.X and PhotonQ), the EU/BMBF via QuantERA (ShoQC), and the Deutsche Forschungsgemeinschaft (DFG, German Research Foundation) – Project-ID 429529648 – TRR 306 QuCoLiMa (“Quantum Cooperativity of Light and Matter”). We acknowledge support from UTokyo Foundation and donations from Nichia Corporation of Japan. W.A. and M.E. acknowledge support from Research Foundation for OptoScience and Technology. A.K. acknowledges financial support from The Forefront Physics and Mathematics Program to Drive Transformation (FoPM). We would like to thank T. Mitani for careful proofreading of the manuscript.

\bmhead{Author contributions}
K.T. conceived the protocol and conducted the simulation. K.T., K.F., A.K., W.A., M.E., J.Y., and P.v.L. formulated the protocol. A.F. supervised the project. K.T. wrote the manuscript with assistance from all other coauthors.

\bmhead{Declarations}
The authors declare no competing interests.

\clearpage


\clearpage

\clearpage

\begin{table}[t]
\begin{center}
\begin{minipage}{274pt}
\caption{{\bf Condition and results of the simulation.} We simulate the generation of the codewords and three kinds of magic states. We numerically perform the coherent bifurcation on squeezed states or seed states $N$ times. In each coherent bifurcation, $n$ photons are detected and we assume $\sqrt{2n}\Delta_1^{-1}=\sqrt{\pi}$, $\Delta_2 =e^{-1}, g=1$, and $\left(\Delta_2^{2}+g^2\Delta_1^2\right)\Delta_3^2 =1$. The value of $\Delta_2$ is $e^{-1}$ in the case of codewords and $e^{-1.16}$ in the case of magic states. Seed-state generation is simulated according to Eq. (\ref{eq:seed_generation}) with $w=\sqrt{\pi}$. The probability shows the total success probability including the generation of seed states.}\label{tab0}%
\begin{tabular}{cccccc}
\toprule
target state & $n$ & $N$ & success & squeezing & fidelity to \\
$\ket{{\rm target}}$ & & & probability & $10\log_{10} \Delta^2$ (dB) & $\hat{X}^N\ket{{\rm target}}$ \\
\midrule
$\ket{\tilde{0}_{\Delta , \Delta}}$ & 6 &  2    & $1.06\times 10^{-3}$   & 6.4  & 0.999  \\
 & 6 &  3    & $5.75\times 10^{-5}$   & 6.9  & 0.996  \\
 & 10 &  3    & $1.65\times 10^{-5}$   & 8.6  & 0.998  \\
 & 10 &  4    & $6.12\times 10^{-7}$   & 8.8  & 0.998  \\
 & 16 &  3    & $5.05\times 10^{-6}$   & 10.3  & 0.998  \\
 & 16 &  4    & $1.18\times 10^{-7}$   & 10.6  & 0.998  \\
\midrule
$\cos{\frac{\pi}{8}} \ket{\tilde{0}_{\Delta , \Delta}} + \sin{\frac{\pi}{8}} \ket{\tilde{1}_{\Delta , \Delta}}$ & 16 & 3    & $2.22\times 10^{-10}$   & 10.2  & 0.997  \\
 & 16 & 4    & $4.97\times 10^{-12}$   & 10.6  & 0.998  \\
\midrule
$\frac{1}{\sqrt{2}} \ket{\tilde{0}_{\Delta , \Delta}} + \frac{e^{-i\frac{\pi}{4}}}{\sqrt{2}} \ket{\tilde{1}_{\Delta , \Delta}}$ & 16 & 3    & $3.53\times 10^{-10}$   & 10.3  & 0.995  \\
 & 16 & 4    & $8.06\times 10^{-12}$   & 10.5  & 0.997  \\
\midrule
$\cos{\theta} \ket{\tilde{0}_{\Delta , \Delta}} + \sin{\theta} \ket{\tilde{1}_{\Delta , \Delta}}\ \ast$ & 16 & 3    & $2.37\times 10^{-10}$   & 10.3  & 0.996  \\
 & 16 & 4    & $5.32\times 10^{-12}$   & 10.4  & 0.997  \\
\botrule
\end{tabular}
\footnotetext[\ast]{$\cos{2\theta}=1/\sqrt{3}$.}
\end{minipage}
\end{center}
\end{table}

\clearpage

\clearpage


\begin{thebibliography}{10}
\providecommand{\url}[1]{{#1}}
\providecommand{\urlprefix}{URL }
\providecommand{\doi}[1]{\url{https://doi.org/#1}}
\bibcommenthead

\bibitem{Gottesman2009}
D.~Gottesman.
\newblock An introduction to quantum error correction and fault-tolerant
  quantum computation.
\newblock \urlprefix\url{https://arxiv.org/abs/0904.2557}

\bibitem{Gottesman1999}
D.~Gottesman, in \emph{Quantum Computing and Quantum Communications}, ed. by
  C.P. Williams (Springer Berlin Heidelberg, Berlin, Heidelberg, 1999), pp.
  302--313

\bibitem{Fowler2012}
A.G. Fowler, M.~Mariantoni, J.M. Martinis, A.N. Cleland, Surface codes: Towards
  practical large-scale quantum computation.
\newblock Phys. Rev. A \textbf{86}, 032324 (2012).
\newblock \urlprefix\url{https://link.aps.org/doi/10.1103/PhysRevA.86.032324}

\bibitem{Preskill2018}
J.~Preskill, Quantum {C}omputing in the {NISQ} era and beyond.
\newblock {Quantum} \textbf{2}, 79 (2018).
\newblock \urlprefix\url{https://doi.org/10.22331/q-2018-08-06-79}

\bibitem{Arute2019}
F. Arute {\it et. al,}, Quantum supremacy using a programmable superconducting processor.
\newblock Nature \textbf{574}, 505--510 (2019).
\newblock \urlprefix\url{https://doi.org/10.1038/s41586-019-1666-5}


\bibitem{Gottesman2001}
D.~Gottesman, A.~Kitaev, J.~Preskill, Encoding a qubit in an oscillator.
\newblock Phys. Rev. A \textbf{64}, 012310 (2001).
\newblock \urlprefix\url{https://link.aps.org/doi/10.1103/PhysRevA.64.012310}

\bibitem{Leghtas2013}
Z.~Leghtas, G.~Kirchmair, B.~Vlastakis, R.J. Schoelkopf, M.H. Devoret,
  M.~Mirrahimi, Hardware-efficient autonomous quantum memory protection.
\newblock Phys. Rev. Lett. \textbf{111}, 120501 (2013).
\newblock
  \urlprefix\url{https://link.aps.org/doi/10.1103/PhysRevLett.111.120501}

\bibitem{Mirrahimi2014}
M.~Mirrahimi, Z.~Leghtas, V.V. Albert, S.~Touzard, R.J. Schoelkopf, L.~Jiang,
  M.H. Devoret, Dynamically protected cat-qubits: a new paradigm for universal
  quantum computation.
\newblock New J. Phys. \textbf{16}, 045014 (2014).
\newblock \urlprefix\url{http://dx.doi.org/10.1088/1367-2630/16/4/045014}

\bibitem{Michael2016}
M.H. Michael, M.~Silveri, R.T. Brierley, V.V. Albert, J.~Salmilehto, L.~Jiang,
  S.M. Girvin, New class of quantum error-correcting codes for a bosonic mode.
\newblock Phys. Rev. X \textbf{6}, 031006 (2016).
\newblock \urlprefix\url{https://link.aps.org/doi/10.1103/PhysRevX.6.031006}

\bibitem{LarsenMikkel2019}
M.V. Larsen, X.~Guo, C.R. Breum, J.S. Neergaard-Nielsen, U.L. Andersen,
  Deterministic generation of a two-dimensional cluster state.
\newblock Science \textbf{366}, 369--372 (2019).
\newblock \urlprefix\url{https://doi.org/10.1126/science.aay4354}

\bibitem{Warit2019}
W.~Asavanant, Y.~Shiozawa, S.~Yokoyama, B.~Charoensombutamon, H.~Emura, R.N.
  Alexander, S.~Takeda, J.~Yoshikawa, N.C. Menicucci, H.~Yonezawa, A.~Furusawa,
  Generation of time-domain-multiplexed two-dimensional cluster state.
\newblock Science \textbf{366}, 373--376 (2019).
\newblock \urlprefix\url{https://doi.org/10.1126/science.aay2645}

\bibitem{Larsen2021}
M.V. Larsen, X.~Guo, C.R. Breum, J.S. Neergaard-Nielsen, U.L. Andersen,
  Deterministic multi-mode gates on a scalable photonic quantum computing
  platform.
\newblock Nat. Phys. \textbf{17}, 1018--1023 (2021).
\newblock \urlprefix\url{https://doi.org/10.1038/s41567-021-01296-y}

\bibitem{Travaglione2002}
B.C. Travaglione, G.J. Milburn, Preparing encoded states in an oscillator.
\newblock Phys. Rev. A \textbf{66}, 052322 (2002).
\newblock \urlprefix\url{https://link.aps.org/doi/10.1103/PhysRevA.66.052322}

\bibitem{Fluehmann2019}
C.~Fl\"uhmann, T.L. Nguyen, M.~Marinelli, V.~Negnevitsky, K.~Mehta, J.P. Home,
  Encoding a qubit in a trapped-ion mechanical oscillator.
\newblock Nature \textbf{566}, 513--517 (2019).
\newblock \urlprefix\url{https://doi.org/10.1038/s41586-019-0960-6}

\bibitem{Campagne-Ibarcq2020}
P.~Campagne-Ibarcq, A.~Eickbusch, S.~Touzard, E.~Zalys-Geller, N.E. Frattini,
  V.V. Sivak, P.~Reinhold, S.~Puri, S.~Shankar, R.J. Schoelkopf, L.~Frunzio,
  M.~Mirrahimi, M.H. Devoret, Quantum error correction of a qubit encoded in
  grid states of an oscillator.
\newblock Nature \textbf{584}, 368--372 (2020).
\newblock \urlprefix\url{https://doi.org/10.1038/s41586-020-2603-3}

\bibitem{Fukui2022a}
K.~Fukui, M.~Endo, W.~Asavanant, A.~Sakaguchi, J. Yoshikawa, A.~Furusawa,
  Generating the Gottesman-Kitaev-Preskill qubit using a cross-Kerr interaction
  between squeezed light and Fock states in optics.
\newblock Phys. Rev. A \textbf{105}, 022436 (2022).
\newblock \urlprefix\url{https://link.aps.org/doi/10.1103/PhysRevA.105.022436}

\bibitem{Harder2016}
G.~Harder, T.J. Bartley, A.E. Lita, S.W. Nam, T.~Gerrits, C.~Silberhorn,
  Single-mode parametric-down-conversion states with 50 photons as a source for
  mesoscopic quantum optics.
\newblock Phys. Rev. Lett. \textbf{116}, 143601 (2016).
\newblock
  \urlprefix\url{https://link.aps.org/doi/10.1103/PhysRevLett.116.143601}

\bibitem{Becerra2015}
F.E. Becerra, J.~Fan, A.~Migdall, Photon number resolution enables quantum
  receiver for realistic coherent optical communications.
\newblock Nat. Photon. \textbf{9}, 48--53 (2015).
\newblock \urlprefix\url{https://doi.org/10.1038/nphoton.2014.280}

\bibitem{Zhong2022}
H.S. Zhong, H.~Wang, Y.H. Deng, M.C. Chen, L.C. Peng, Y.H. Luo, J.~Qin, D.~Wu,
  X.~Ding, Y.~Hu, P.~Hu, X.Y. Yang, W.J. Zhang, H.~Li, Y.~Li, X.~Jiang, L.~Gan,
  G.~Yang, L.~You, Z.~Wang, L.~Li, N.L. Liu, C.Y. Lu, J.W. Pan, Quantum
  computational advantage using photons.
\newblock Science \textbf{370}, 1460--1463 (2022).
\newblock \urlprefix\url{https://doi.org/10.1126/science.abe8770}

\bibitem{Madsen2022}
L.S. Madsen, F.~Laudenbach, M.F. Askarani, F.~Rortais, T.~Vincent, J.F.F.
  Bulmer, F.M. Miatto, L.~Neuhaus, L.G. Helt, M.J. Collins, A.E. Lita,
  T.~Gerrits, S.W. Nam, V.D. Vaidya, M.~Menotti, I.~Dhand, Z.~Vernon,
  N.~Quesada, J.~Lavoie, Quantum computational advantage with a programmable
  photonic processor.
\newblock Nature \textbf{606}, 75--81 (2020).
\newblock \urlprefix\url{https://doi.org/10.1038/s41586-022-04725-x}

\bibitem{Bravyi2005}
S.~Bravyi, A.~Kitaev, Universal quantum computation with ideal Clifford gates
  and noisy ancillas.
\newblock Phys. Rev. A \textbf{71}, 022316 (2005).
\newblock \urlprefix\url{https://link.aps.org/doi/10.1103/PhysRevA.71.022316}

\bibitem{Knill2005}
E.~Knill, Quantum computing with realistically noisy devices.
\newblock Nature \textbf{434}, 39--44 (2005).
\newblock \urlprefix\url{https://doi.org/10.1038/nature03350}

\bibitem{Vasconcelos2010}
H.M. Vasconcelos, L.~Sanz, S.~Glancy, All-optical generation of states for
  “{E}ncoding a qubit in an oscillator”.
\newblock Opt. Lett. \textbf{35}, 3261--3263 (2010).
\newblock
  \urlprefix\url{http://www.osapublishing.org/ol/abstract.cfm?URI=ol-35-19-3261}

\bibitem{Weigand2018}
D.J. Weigand, B.M. Terhal, Generating grid states from {S}chr\"odinger-cat
  states without postselection.
\newblock Phys. Rev. A \textbf{97}, 022341 (2018).
\newblock \urlprefix\url{https://link.aps.org/doi/10.1103/PhysRevA.97.022341}

\bibitem{Su2019}
D.~Su, C.R. Myers, K.K. Sabapathy, Conversion of Gaussian states to
  non-Gaussian states using photon-number-resolving detectors.
\newblock Phys. Rev. A \textbf{100}, 052301 (2019).
\newblock \urlprefix\url{https://link.aps.org/doi/10.1103/PhysRevA.100.052301}

\bibitem{Sabapathy2019a}
K.K. Sabapathy, H.~Qi, J.~Izaac, C.~Weedbrook, Production of photonic universal
  quantum gates enhanced by machine learning.
\newblock Phys. Rev. A \textbf{100}, 012326 (2019).
\newblock \urlprefix\url{https://link.aps.org/doi/10.1103/PhysRevA.100.012326}

\bibitem{Tzitrin2020}
I.~Tzitrin, J.E. Bourassa, N.C. Menicucci, K.K. Sabapathy, Progress towards
  practical qubit computation using approximate {G}ottesman-{K}itaev-{P}reskill
  codes.
\newblock Phys. Rev. A \textbf{101}, 032315 (2020).
\newblock \urlprefix\url{https://link.aps.org/doi/10.1103/PhysRevA.101.032315}

\bibitem{Fukui2022}
K.~Fukui, S.~Takeda, M.~Endo, W.~Asavanant, J. Yoshikawa, P.~van Loock,
  A.~Furusawa, Efficient backcasting search for optical quantum state
  synthesis.
\newblock Phys. Rev. Lett. \textbf{128}, 240503 (2022).
\newblock
  \urlprefix\url{https://link.aps.org/doi/10.1103/PhysRevLett.128.240503}

\bibitem{Hamilton2017}
C.S. Hamilton, R.~Kruse, L.~Sansoni, S.~Barkhofen, C.~Silberhorn, I.~Jex,
  Gaussian Boson Sampling.
\newblock Phys. Rev. Lett. \textbf{119}, 170501 (2017).
\newblock
  \urlprefix\url{https://link.aps.org/doi/10.1103/PhysRevLett.119.170501}

\bibitem{Takase2021}
K.~Takase, J.~Yoshikawa, W.~Asavanant, M.~Endo, A.~Furusawa, Generation of
  optical {S}chr\"odinger cat states by generalized photon subtraction.
\newblock Phys. Rev. A \textbf{103}, 013710 (2021).
\newblock \urlprefix\url{https://link.aps.org/doi/10.1103/PhysRevA.103.013710}

\bibitem{Dakna1997}
M.~Dakna, T.~Anhut, T.~Opatrn\'y, L.~Kn\"oll, D.G. Welsch, Generating
  {S}chr\"odinger-cat-like states by means of conditional measurements on a
  beam splitter.
\newblock Phys. Rev. A \textbf{55}, 3184--3194 (1997).
\newblock \urlprefix\url{https://link.aps.org/doi/10.1103/PhysRevA.55.3184}

\bibitem{Filip2005}
R.~Filip, P.~Marek, U.L. Andersen, Measurement-induced continuous-variable
  quantum interactions.
\newblock Phys. Rev. A \textbf{71}, 042308 (2005).
\newblock \urlprefix\url{https://link.aps.org/doi/10.1103/PhysRevA.71.042308}

\bibitem{Shiozawa2018}
Y.~Shiozawa, J.~Yoshikawa, S.~Yokoyama, T.~Kaji, K.~Makino, T.~Serikawa,
  R.~Nakamura, S.~Suzuki, S.~Yamazaki, W.~Asavanant, S.~Takeda, P.~van Loock,
  A.~Furusawa, Quantum nondemolition gate operations and measurements in real
  time on fluctuating signals.
\newblock Phys. Rev. A \textbf{98}, 052311 (2018).
\newblock \urlprefix\url{https://link.aps.org/doi/10.1103/PhysRevA.98.052311}

\bibitem{Braunstein2005}
S.L. Braunstein, Squeezing as an irreducible resource.
\newblock Phys. Rev. A \textbf{71}, 055801 (2005).
\newblock \urlprefix\url{https://link.aps.org/doi/10.1103/PhysRevA.71.055801}

\bibitem{Reck1994}
M.~Reck, A.~Zeilinger, H.J. Bernstein, P.~Bertani, Experimental realization of
  any discrete unitary operator.
\newblock Phys. Rev. Lett. \textbf{73}, 58--61 (1994).
\newblock \urlprefix\url{https://link.aps.org/doi/10.1103/PhysRevLett.73.58}

\bibitem{Loock2007}
P.~van Loock, C.~Weedbrook, M.~Gu, Building Gaussian cluster states by linear
  optics.
\newblock Phys. Rev. A \textbf{76}, 032321 (2007).
\newblock \urlprefix\url{https://link.aps.org/doi/10.1103/PhysRevA.76.032321}

\bibitem{Fukui2018}
K.~Fukui, A.~Tomita, A.~Okamoto, K.~Fujii, High-threshold fault-tolerant
  quantum computation with analog quantum error correction.
\newblock Phys. Rev. X \textbf{8}, 021054 (2018).
\newblock \urlprefix\url{https://link.aps.org/doi/10.1103/PhysRevX.8.021054}

\bibitem{Takase2022a}
K.~Takase, A.~Kawasaki, B.K. Jeong, T.~Kashiwazaki, T.~Kazama, K.~Enbutsu,
  K.~Watanabe, T.~Umeki, S.~Miki, H.~Terai, M.~Yabuno, F.~China, W.~Asavanant,
  M.~Endo, J. Yoshikawa, A.~Furusawa, Quantum arbitrary waveform generator.
\newblock Sci. Adv. \textbf{8}, eadd4019 (2022).
\newblock \urlprefix\url{https://doi.org/10.1126/sciadv.add4019}

\bibitem{Walshe2020}
B.W. Walshe, B.Q. Baragiola, R.N. Alexander, N.C. Menicucci,
  Continuous-variable gate teleportation and bosonic-code error correction.
\newblock Phys. Rev. A \textbf{102}, 062411 (2020).
\newblock \urlprefix\url{https://link.aps.org/doi/10.1103/PhysRevA.102.062411}

\bibitem{Baragiola2019}
B.Q. Baragiola, G. Pantaleoni, R.N. Alexander, A. Karanjai, N.C. Menicucci, All-Gaussian Universality and Fault Tolerance with the {G}ottesman-{K}itaev-{P}reskill Code.
\newblock Phys. Rev. Lett. \textbf{123}, 200502 (2019).
\newblock \urlprefix\url{https://link.aps.org/doi/10.1103/PhysRevLett.123.200502}

\bibitem{Yamasaki2020}
H. Yamasaki, T. Matsuura, M. Koashi, Cost-reduced all-Gaussian universality with the Gottesman-Kitaev-Preskill code: Resource-theoretic approach to cost analysis.
\newblock Phys. Rev. Res. \textbf{2}, 023270 (2020).
\newblock \urlprefix\url{https://link.aps.org/doi/10.1103/PhysRevResearch.2.023270}

\bibitem{Konno2021}
S.~Konno, W.~Asavanant, K.~Fukui, A.~Sakaguchi, F.~Hanamura, P.~Marek,
  R.~Filip, J. Yoshikawa, A.~Furusawa, Non-Clifford gate on optical qubits by
  nonlinear feedforward.
\newblock Phys. Rev. Res. \textbf{3}, 043026 (2021).
\newblock
  \urlprefix\url{https://link.aps.org/doi/10.1103/PhysRevResearch.3.043026}

\bibitem{Sakaguchi2022}
A.~Sakaguchi, S.~Konno, F.~Hanamura, W.~Asavanant, K.~Takase, H.~Ogawa,
  P.~Marek, R.~Filip, J. Yoshikawa, E.~Huntington, H.~Yonezawa, A.~Furusawa.
\newblock Nonlinear feedforward enabling quantum computation.
\newblock \urlprefix\url{https://arxiv.org/abs/2210.17120}

\bibitem{Kashiwazaki2021b}
T.~Kashiwazaki, T.~Yamashima, N.~Takanashi, A.~Inoue, T.~Umeki, A.~Furusawa,
  Fabrication of low-loss quasi-single-mode {PPLN} waveguide and its
  application to a modularized broadband high-level squeezer.
\newblock Appl. Phys. Lett. \textbf{119}, 251104 (2021).
\newblock \urlprefix\url{https://doi.org/10.1063/5.0063118}

\bibitem{Rancic2018}
M.~Ran\v{c}i\'{c}, M.P. Hedges, R.L. Ahlefeldt, M.J. Sellars, Coherence
  time of over a second in a telecom-compatible quantum memory
  storage material.
\newblock Nat. Phys. \textbf{14}, 50--54 (2018).
\newblock \urlprefix\url{https://doi.org/10.1038/nphys4254}

\bibitem{Joshi2018}
C.~Joshi, A.~Farsi, S.~Clemmen, S.~Ramelow, A.L. Gaeta, Frequency multiplexing
  for quasi-deterministic heralded single-photon sources.
\newblock Nat. Commun. \textbf{9}, 847 (2018).
\newblock \urlprefix\url{https://doi.org/10.1038/s41467-018-03254-4}

\bibitem{Fukui2019arxiv}
K.~Fukui.
\newblock High-threshold fault-tolerant quantum computation with the GKP qubit
  and realistically noisy devices.
\newblock \urlprefix\url{https://arxiv.org/abs/1906.09767}

\end{thebibliography}
\end{document}